\documentclass[aps,superscriptaddress,showpacs,nofootinbib,eqsecnum]{revtex4}

\usepackage{amsmath,lscape,epsfig} \usepackage{amsfonts} \usepackage{amsmath}
\usepackage{amssymb} \usepackage{graphicx} \usepackage{multirow}

\usepackage{slashed}
\usepackage{amsmath,lscape,epsfig}
\usepackage{amsfonts}

\usepackage{amsmath}
\usepackage{amssymb}
\usepackage{graphicx,marvosym}

\usepackage{amsmath,lscape,epsfig} \usepackage{amsfonts}
\usepackage{amsmath} \usepackage{amssymb} \usepackage{graphicx}

\newcommand{\tauv}{\boldsymbol{\tau}}

\newcommand{\piv}{\boldsymbol{\pi}}
\newcommand{\bfr}{\boldsymbol{r}}			
\newcommand{\MeV}{{\rm MeV}}				
\newcommand{\GeV}{{\rm GeV}}				
\newcommand{\fm}{{\rm fm}}				

\def\i{{\rm i}}
\def\uq{{\rm u}}
\def\dq{{\rm d}}

\newcommand{\SKIP}[1]{$\!\!$}			


\begin{document}

\title{The Surface Tension of Quark Matter in a Geometrical Approach} 

\author{Marcus B.\ Pinto} 
\affiliation{Departamento de Fisica, Universidade Federal de
  Santa Catarina, 88040-900 Florian\'{o}polis, Santa Catarina, Brazil} 

\author{Volker Koch} \affiliation{Nuclear Science Division,
  Lawrence Berkeley National Laboratory, Berkeley, CA 94720, USA}

\author{J{\o}rgen Randrup} \affiliation{Nuclear Science Division,
  Lawrence Berkeley National Laboratory, Berkeley, CA 94720, USA}

\date{\today}

\begin{abstract}
The surface tension of quark matter plays a crucial role
for the possibility of quark matter nucleation
during the formation of compact stellar objects,
because it determines the nucleation rate and the associated critical size.
However, this quantity is not well known and the theoretical estimates
fall within a wide range,  $\gamma_0\approx5-300{\rm\ MeV/fm^2}$.  
We show here that once the equation of state is available 
one may use a geometrical approach to obtain 
a numerical value for the surface tension
that is consistent with the model approximations adopted.  
We illustrate this method within  the two-flavor linear $\sigma$ model
and the Nambu--Jona-Lasinio model with two and three flavors.
Treating these models in the mean-field approximation,
we find  $\gamma_0 \approx7-30{\rm\ MeV/fm^2}$. 
Such a relatively small surface tension 
would favor the formation of quark stars 
and may thus have significant astrophysical implications. 
We also investigate how the surface tension decreases towards zero
as the temperature is raised from zero to its critical value.
\end{abstract}

\maketitle

\section{Introduction}

Lattice-gauge calculations yield a non-vanishing value of the quark condensate
$\langle{\overline \psi} \psi \rangle$ in the QCD vacuum \cite{Aoki},
indicating that chiral symmetry is broken.
This general feature of the vacuum remains present even for massless quarks
because the symmetry is then broken spontaneously.
On the other hand, chiral symmetry is expected to become restored 
at sufficiently high values of the net-baryon density $\rho$
or/and the temperature $T$.
The character of this phase change is not yet well understood
but it has significant implications in areas 
such as cosmology and astrophysics
and it is a focal point for current experimental and theoretical research
in nuclear physics.

Nuclear collision experiments carried out with
the Relativistic Heavy Ion Collider (RHIC) at 
the Brookhaven National Laboratory
and with the Large Hadron Collider at CERN
explore systems having relatively small net-baryon densities $\rho$
and the associated chemical potentials $\mu$ are negligible.
Lattice calculations can readily be carried out at vanishing $\mu$
and they indicate that a cross-over transformation 
from the chirally broken phase to the restored phase occurs
as the temperature is increased from below to above 
the cross-over temperature $T_\times\approx160\,{\rm MeV}$ 
\cite{Aoki,Aoki:2006br,Bazavov:2011nk}.

The other extreme region of the QCD phase diagram,
namely low temperatures and high chemical potentials,
cannot be addressed by current lattice-QCD methods,
due to the fermion sign problem,
and studies of this phase region 
must therefore rely on less fundamental models.
Most investigations suggest that there is a first-order phase transition 
which, for $T\approx0$, sets in at baryon densities several times that of the
nuclear saturation density, $\rho_0\approx0.153/\fm^3$. 
The properties of strongly interacting matter in this phase region
are important for our understanding of compact stars.

If indeed such a first-order phase transition exists at $T=0$,
then, as the temperature is raised,
one would expect it to remain present 
but gradually weaken and eventually terminate
at a critical point $(\mu_c,T_c)$.
The existence and location of such a critical point is a subject of
intense theoretical investigation with a variety of models,
including in particular effective-field models,
such as the linear $\sigma$ model (LSM),
and effective quark models,
such as the Nambu--Jona-Lasinio (NJL),
at different levels of sophistication considering up to three quark flavors
and possibly including the Polyakov loop to account for confinement 
\cite {pedropoly,mao}. 
Experimentally the corresponding region of density and excitation
may be produced in current nuclear collisions at the low-energy end of RHIC 
and in the future with FAIR at GSI and NICA at JINR
which are being constructed with such investigations in mind.

In the present work, we concentrate on the high-$\mu$ and low-$T$
part of the phase diagram with the aim of exploring
the expected chiral phase transition
which has significant implications for the possible existence of
quark stars \cite {ed1,bie}. 
It should be noted that chiral symmetry may be restored already during 
the early post-bounce accretion stage of a core-collapse supernova event 
and the associated neutrino burst might then provide a spectacular signature 
for the presence of quark matter inside compact stars \cite {sagert}.
However, as pointed out in Refs.\ \cite{bruno,bombaci},
the possibilities depend on the dynamics of the phase conversion 
and especially on the time scales involved.

When the phase diagram of bulk matter exhibits a first-order phase transition,
the two phases may coexist in mutual thermodynamic equilibrium and,
consequently, when brought into physical contact
a mechanically stable interface will develop between them.
The associated interface tension $\gamma_T$
(which we shall often refer to simply as the surface tension of quark matter)
depends on the temperature $T$; it has its largest magnitude at $T=0$
and approaches zero as $T$ is increased to $T_c$.
This quantity
plays a key role in the phase conversion process
and it is related to various characteristic quantities such as 
the nucleation rate, the critical bubble radius, and the favored scale of the
blobs generated by the spinodal instabilities \cite{PhysRep,RandrupPRC79}.
(As we shall see, the surface tension is essentially proportional to the
effective interaction range, which determines the width of the surface region,
and the spatial size of the most rapidly amplified density irregularity
is also proportional to this quantity.)

Unfortunately, despite its central importance, 
the surface tension of quark matter is rather poorly known.
Estimates in the literature fall within a wide range,
typically $\gamma_0\approx10-50~\MeV/\fm^2$ 
\cite{heiselberg,sato} and values of $\gamma_0\approx30\,\MeV/\fm^2$ 
have been considered for studying the effect of quark matter 
nucleation on the evolution of proto-neutron stars \cite{constanca}.
But the authors in Ref.\ \cite {russo}, taking into
account the effects from charge screening and structured mixed phases,
estimate $\gamma_0\approx50-150~\MeV/\fm^2$, without excluding smaller values,
and an ever higher value, $\gamma_0\approx300~\MeV/\fm^2$, 
was found by Alford {\em et al.}\ \cite {alford} 
on the basis of dimensional analysis of the minimal
interface between a color-flavor locked phase and nuclear matter.

The surface tension for two-flavor quark matter was evaluated within
the framework of the LSM by Palhares and Fraga \cite{leticia}. 
In that work the authors considered the one-loop effective potential 
and then fitted its relevant part, 
which included both the chirally symmetric and broken state, 
by a quartic polynomial. 
The surface tension was evaluated
using the thin-wall approximation for bubble nucleation and the
estimated values cover the  $5-15\,\MeV/\fm^2$ range, 
depending on the inclusion of vacuum and/or thermal corrections. 
In principle, this range makes nucleation of quark matter possible 
during the early post-bounce stage of core-collapse supernovae
and it is thus a rather important result. 
It is also worth noting that a small surface tension would 
facilitate various structures in compact stars,
including mixed phases \cite {kurkela}.

The present work is devoted to the evaluation of the surface tension
for  quark matter using both the LSM (with two flavors) and the
NJL model (with two and three flavors) following the procedure employed
in Ref.\ \cite {RandrupPRC79}. Here, the LSM is mainly included to check the
consistency of our procedure by comparing our present results with those
obtained by the thin-wall approximation of Ref.\ \cite{leticia}
(we find the agreement to be very good).
The NJL model is considered with two and three flavors
because the latter, which contains strangeness, is  one of the most popular 
effective quark models used in studies related to compact stars.
As explained below,
the method described in Ref.\ \cite{RandrupPRC79} makes it possible
to express the surface tension for any subcritical temperature in terms of 
the free energy density for uniform matter in the unstable density range.
Because the models employed readily provide the equation of state (EoS)
for the full density range, they are well suited for our purpose
and we may directly employ the method without any further approximations.
In practice, the  procedure is rather simple to implement 
and it provides an estimate for the surface tension
that is consistent with the EoS implied by the model employed,
with its specific approximations and parametrizations. 

The paper is organized as follows. 
In Sect.\ \ref{method} we review the method for extracting the surface tension
from the equation of state. 
In Sec.\ \ref{models} we then present the two-flavor versions of the two models
considered and discuss how to extract the surface tension.
Section \ref{NJL3} is devoted to the treatment
for the more realistic $SU(3)$ version of the NJL model
and our numerical results are presented in Sec.\ \ref{results},
both for cold matter and for temperatures up to the critical value.
The conclusions and final remarks are presented in Sec.\ \ref{conclude}.

\section{The Geometric Approach to the Surface Tension Evaluation }
\label{method}

We assume here that the material at hand, strongly interacting matter,
may appear in two different phases under the same thermodynamic conditions
of temperature $T$, chemical potential $\mu$, and pressure $P$.
These two coexisting phases have different values of other relevant quantities,
such as the energy density $\cal E$, the (net baryon) density $\rho$,
and the entropy density $s$.
Under such circumstances, 
the two phases will develop a mechanically stable interface if placed
in physical contact and it is the purpose of the present study
to evaluate the associated interface tension,  $\gamma_T$.

The two-phase feature appears for all temperatures below the
critical value, $T_c$.  Thus, for any subcritical temperature, $T<T_c$,
hadronic matter at the (net-baryon) density $\rho_1(T)$ 
has the same chemical potential and pressure
as quark matter at the (larger) density $\rho_2(T)$.
As $T$ is increased from zero to $T_c$,
the coexistence phase points $(\rho_1,T)$ and $(\rho_2,T)$ trace out the 
lower and higher branches of the phase coexistence boundary, respectively,
gradually approaching each other and finally coinciding for $T=T_c$.
Any $(\rho,T)$ phase point outside of this boundary corresponds to
thermodynamically stable uniform matter,
whereas uniform matter prepared with a density and temperature corresponding
to a phase point inside the phase coexistence boundary
is thermodynamically unstable and prefers to separate into
two coexisting thermodynamically stable phases
separated by a mechanically stable interface.
Because such a two-phase configuration is in global thermodynamic equilibrium, 
the local values of $T$, $\mu$, and $P$ remain unchanged
as one moves from the interior of one phase through the interface
region and into the interior of the partner phase,
as the local density $\rho$ increases steadily from 
the lower coexistence value $\rho_1$ 
to the corresponding higher coexistence value $\rho_2$.

It is convenient to work in the canonical framework in which
where the control parameters are temperature and density.
The basic thermodynamic function is thus $f_T(\rho)$,
the free energy density as a function of the (net baryon) density $\rho$
for the specified temperature $T$.
The chemical potential can then be recovered as 
$\mu_T(\rho)=\partial_\rho f_T(\rho)$,
and the entropy density as $s_T(\rho)=-\partial_T f_T(\rho)$,
so the energy density is ${\cal E}_T(\rho)=f_T(\rho)-T\partial_Tf_T(\rho)$,
while the pressure is $P_T(\rho)=\rho\partial_\rho f_T(\rho)-f_T(\rho)$.

For single-phase systems $f_T(\rho)$ is convex,
{\em i.e.}\ its second derivative $\partial_\rho^2f_T(\rho)$ is positive,
 while the appearance of a concavity in $f_T(\rho)$ 
signals the occurrence of phase coexistence, at that temperature.
This is easily understood because when $f_T(\rho)$ has a local concave
anomaly then there exist a pair of densities, $\rho_1$ and $\rho_2$,
for which the tangents to $f_T(\rho)$ are common.
Therefore $f_T(\rho)$ has the same slope at those two densities,
so the corresponding chemical potentials are equal,
$\mu_T(\rho_1)=\partial_\rho f_T(\rho_1)
	=\partial_\rho f_T(\rho_2)=\mu_T(\rho_2)$.
Furthermore, because a linear extrapolation of $f_T(\rho)$ leads from 
one of the touching points to the other, also the two pressures are equal,
$P_T(\rho_1)=\rho_1\partial_\rho f_T(\rho_1)-f_T(\rho_1)
	=\rho_2\partial_\rho f_T(\rho_2)-f_T(\rho_2)=P_T(\rho_2)$.
So uniform matter at the density $\rho_1$ has the same temperature,
chemical potential, and pressure as uniform matter at the density $\rho_2$.
The common tangent between the two coexistence points
corresponds to the familiar Maxwell construction and shall here 
be denoted as $f_T^M(\rho)$.
Obviously, $f_T(\rho)$ and $f_T^M(\rho)$ coincide at the two coexistence
densities and, furthermore, 
$f_T(\rho)$ exceeds $f_T^M(\rho)$ for intermediate densities.
Therefore we have $\Delta f_T(\rho)\equiv f_T(\rho)-f_T^M(\rho)\geq0$.

For a given (subcritical) temperature $T$,
we now consider a configuration in which the two coexisting bulk phases
are placed in physical contact along a planar interface.
The associated equilibrium profile density is denoted by $\rho_T(z)$ 
where $z$ denotes the location in the direction normal to the interface.
In the diffuse interface region, the corresponding local free energy density,
$f_T(z)$, differs from what it would be for the corresponding Maxwell system,
{\em i.e.}\ a mathematical mix of the two coexisting bulk phases
with the mixing ratio adjusted to yield an average density 
equal to the local value $\rho(z)$.
This local deficit amounts to
\begin{equation}
\delta f_T(z)\ =\  f_T(z)-f_i
	-\frac{f_T(\rho_2)-f_T(\rho_1)}{\rho_2-\rho_1}\,(\rho_T(z)-\rho_i)\ ,
\end{equation}
where $\rho_i$ is either one of the two coexistence densities.
The function $\delta f_T(z)$ is smooth and it tends quickly to zero
away from the interface where $\rho_T(z)$ rapidly approaches $\rho_i$
and $f_T(z)$ rapidly approaches $f_T(\rho_i)$.
The interface tension $\gamma_T$ is the total deficit in free energy
per unit area of planar interface,
\begin{equation}
\gamma_T\ =\ \int_{-\infty}^{+\infty}\delta f_T(z)\,dz\ .
\end{equation}

As discussed in Ref.\ \cite{RandrupPRC79},
when a gradient term used to take account of finite-range effects,
the tension associated with the interface between the two phases
can be expressed without explicit knowledge about the profile functions
but exclusively in terms of the equation of state for uniform
(albeit unstable) matter,
\begin{equation}\label{gamma}
 \gamma_T\ =\ a \int_{\rho_1(T)}^{\rho_2(T)}
\left[ 2 {\cal E}_g \Delta f_T(\rho)\right]^{1/2} \frac {d \rho}{\rho_g}\ ,
\end{equation}
where $\rho_g$ is a characteristic value of the density
and ${\cal E}_g$ is a characteristic value of the energy density,
while the parameter $a$ is an effective interaction range related to
the strength of the gradient term, $C=a^2 {\cal E}_g/\rho_g^2$.
We choose the characteristic phase point to be in the middle of the 
coexistence region, $\rho_g=\rho_c$ and
${\cal E}_g=[{\cal E}_0(\rho_c)+{\cal E}_c]/2$,
where ${\cal E}_0(\rho_c)$ is energy density at $(\rho_c,T=0)$,
while ${\cal E}_c$ is energy density at the critical point $(\rho_c,T_c)$. 
The length $a$ as a somewhat adjustable parameter
governing the width of the interface region 
and the magnitude of the tension \cite{RandrupPRC79}.
For the LSM it is natural to expect that $a\approx1/m_\sigma\approx0.33\,\fm$ 
which, also, 
is approximately the value found in an application of the Thomas-Fermi 
approximation to the NJL model  \cite {russos}.  
Therefore, we shall adopt the value $a=0.33\,{\rm fm}$ 
throughout the present work.
While there is some arbitrariness in fixing these quantities, 
it is reassuring that the resulting surface tension is in excellent agreement 
with the value obtained in Ref.\ \cite{leticia}.

With these parameters fixed,
the interface tension can be calculated once the free energy density
$f_T(\rho)$ is known for uniform matter in the unstable phase region,
$\rho_1(T)\leq\rho\leq\rho_2(T)$.
While this is straightforward in a canonical formulation,
where each $(\rho,T)$ characterizes only one manifestation of the system,
even inside the unstable phase region,
the task is more complicated in the commonly used grand canonical formulation 
because a given $(\mu,T)$ phase point characterizes three different 
manifestations of the system, 
one stable, one metastable, and one unstable.
The metastable solutions are located near the coexistence densities,
while the unstable solutions are located in the intermediate spinodal region
where uniform matter is mechanically unstable so that even infinitesimal
irregularities may be exponentially amplified.
By contrast, only irregularities of a sufficient amplitude are amplified
in the metastable regions, leading towards either nucleation
(near the lower coexistence density $\rho_1$)
or bubble formation (near the higher coexistence density $\rho_2$).

\section{The EoS for the effective two flavor quark models}
\label{models}

In this section, 
we review the mean-field  results for the thermodynamic potential 
for the two effective models when only two quark flavors are included.
These results have been widely discussed in the literature and here we
follow Ref.\ \cite {scavenius} 
(see Ref.\ \cite {prc1} for results beyond the mean-field approximation). 
The two models are similar in the sense that they do not have confinement and 
they incorporate spontaneous chiral symmetry breaking,
which happens at the classical level in the LSM 
but only via quantum corrections in the NJL model. 
The fermionic fields representing the quarks
are the only degrees of freedom in the NJL model at the tree level,
while the LSM also contains scalar ($\sigma$) and pseudoscalar ($\piv$)
meson fields.

\subsection {The linear $\sigma$ model}

In standard notation, the Lagrangian density of the LSM with quarks reads
\begin{equation}
\mathcal{L}_{\rm LSM}\ =\ \frac{1}{2}\left(\partial_{\mu} \piv \right)^2+
\frac{1}{2}\left(\partial_{\mu} \sigma \right)^2-U\left(\sigma,\piv\right)
+{\bar \psi}\left[\i \gamma^\mu \partial_{\mu} 
	-g\left(\sigma + \i \gamma_5 \tauv \cdot \piv \right)\right]\psi\; ,
\label{lsm1}
\end{equation}
where $\psi$ is the flavor isodoublet spinor representing the quarks 
($\uq$ and $\dq$), and 
\begin{equation}
U\left( \sigma,\piv \right)\
=\ \frac{\lambda^2}{4}\left(\sigma^2+\piv^2-v^2\right)^2-H\sigma \; ,
\label{lsm2}
\end{equation}
is the classical potential energy density. 
In the chiral limit (obtained for $H=0$) the chiral symmetry, 
$SU(2)_V\times SU(2)_A$, is spontaneously broken at the classical level, 
and the pion is the associated massless Goldstone boson.   
For $H \neq 0$, the chiral symmetry is explicitly broken by the last term
 in $U( \sigma,\piv)$ which gives the pion a finite mass at vanishing
$T$ and $\mu$. 
The parameters are usually chosen so that chiral symmetry is spontaneously
broken in the vacuum and the expectation values of the meson fields
are $\langle \sigma \rangle=f_\pi$ and $\langle \piv \rangle={\boldsymbol{0}}$
 where $f_\pi=93 \, {\rm MeV}$ is the pion decay constant. 
Following Ref.\ \cite {scavenius}, we fix the parameters as follows: 
$v^2\simeq(87.73\,\MeV)^2$, $\lambda^2 \simeq 20$, 
and $H \simeq (12.1\,\GeV)^3$.
Using the standard relations, 
$H=f_\pi m_\pi^2$, $v^2= f_\pi^2-m_\pi^2/\lambda^2$,
and $m_\sigma^2=2\lambda^2f_\pi^2$,
we obtain the meson masses, $m_\pi=138\,\MeV$ and $m_\sigma=600\,\MeV$.
The coupling constant $g$ is usually fixed so that the effective quark mass 
in vacuum, $M^{\rm vac}=g f_\pi$, be about one third of the nucleon mass, 
which gives $g \simeq 3.3$. We note that the
same parameter set was also used to evaluate the surface tension,
$\gamma_T$, in Ref. \cite{leticia}.
To the one-loop level, the grand canonical potential is obtained 
by integrating the action over the fermionic fields \cite {scavenius},
\begin{equation}\label{omegalsm}
 \Omega_{\rm LSM}(\sigma,\piv;T,\mu_q) = 
U\left(\sigma,\piv \right)-2N_f N_c 
\int\frac{d^3\boldsymbol{p}}{(2\pi)^3}
\left\{ E-T\ln\left[ 1- n^+ \right] -T\ln \left[ 1- n^-\right ]\right\}\ ,
\end{equation}
where $N_c=3$, $N_f=2$,  $E^{2}=\boldsymbol{p}^2+M^{2}$,  and
$n^\pm = \{ 1+\exp[(E\mp\mu_q)/T]\}^{-1}$ represent the particle/antiparticle 
thermal occupancies with $\mu_q=\mu_u=\mu_d$, where $\mu= 3 \mu_q$.
For for given values of $T$ and $\mu$,
the equilibrium values of the meson fields are obtained by minimizing 
$\Omega(\sigma,\piv;T,\mu_q)$ with respect to those,
yielding the most likely values $\underline{\sigma}$ and $\underline{\piv}$.
The latter one vanishes in the mean-field approximation, so the associated 
constituent quark mass is given as $M^2=g^2\underline{\sigma}^2$.
The minimum value of the grand potential represents minus
the equilibrium pressure, $\Omega_{\rm min}(T,\mu)=-P$,
so the net quark density is given by $\rho_q=(\partial P/\partial\mu_q)_T$, 
where $\rho_q=3\rho$. 
The entropy density given by $s=(\partial P/\partial T)_{\mu_q}$,
while the energy density, $\cal E$, can then be obtained by means of 
the standard thermodynamic relation $P=Ts-{\cal E}+\mu \rho$
and the free energy density is $f\equiv{\cal E}-Ts=\mu\rho-P$.

In the neighborhood of the phase coexistence line in the $(\mu,T)$ plane, 
the grand potential has three extrema representing 
stable, metastable, and spinodally unstable matter.
As emphasized above, the extraction of  the surface tension by the
geometric approach requires the consideration of all three extrema.

In contrast to the NJL model, the vacuum term represented
by the first term in the integrand of Eq.\ (\ref {omegalsm}) 
is not essential for the spontaneous breaking of chiral symmetry.
In the LSM this already happens at the classical level
and the symmetry restoration is driven mainly by the terms containing $n^\pm$.
Therefore, we neglect the vacuum term in the present LSM application,
where the aim is to compare our estimates with the zero-temperature
interface tension obtained in Ref.\ \cite{leticia},
$\gamma_0\simeq12.98\,\MeV/\fm^2$, where the relevant part of 
the same thermodynamical potential was fitted with a quartic polynomial. 
In our approach such a fitting procedure is not necessary because 
the thermodynamic potential is evaluated for all values of $\mu$ and $T$. 
This will lead to somewhat different numerical values for the surface tension
$\gamma_T$. 
It was shown in Ref.\ \cite{leticia} that the inclusion of vacuum terms 
at $T=0$ increases the surface tension value from
$\gamma_0= 12.98\,\MeV/\fm^2$ to about
$\gamma_0\simeq17\,\MeV/\fm^2$. 
In practice, further refinements including vacuum and in-medium two-loop
corrections are possible by following the same technical steps that
were employed in the evaluation of the thermodynamical potential for
the Yukawa theory \cite {yukawa}. 

It is now straightforward to determine the phase-coexistence line 
in the $(\mu,T)$ plane which forms the starting point 
for determining all quantities related to $\gamma_T$. 
It starts at $(\mu=918\,\MeV,T=0)$ and terminates
at the critical point $(\mu_c=621\,\MeV,T_c=99\,\MeV)$,
which agrees with Ref.\ \cite {scavenius}.  
We now have all the ingredients needed for determining 
the coexistence densities $\rho_1(T)$ and $\rho_2(T)$
as well as the characteristic values $\rho_g$ and ${\cal E}_g$ 
appearing in Eq.\ (\ref {gamma}).

The difference $\Delta f(\rho)$ can be readily determined numerically 
by considering the stable (global) minimum, the metastable (local) minimum, 
and the unstable (local) maximum appearing in the thermodynamical potential, 
as will be explicitely shown in section \ref{results}.
Although our method for extracting the surface tension
does not require the profiles $\sigma(z)$ and $\rho(z)$, these functions 
do provide interesting additional information about the interface.
To obtain the profile functions within the LSM, 
it suffices to consider the grand canonical potential 
in the $\sigma$ direction only, {\em i.e.}\ taking $\piv=\boldsymbol{0}$.
At $T=0$ it can be expressed in terms of the Fermi momentum $p_F$ 
(given by $p_F^2=\mu_q^2-g^2\sigma^2$),
\begin{equation}
\Omega_{\rm LSM} (\sigma,\piv=\boldsymbol{0};T=0,\mu)\ =\
U(\sigma) - \frac{N_f N_c}{24 \pi^2} \left \{ 2 \mu_q p_F^3 - 3(g \sigma)^2 
\left[ \mu_q p_F - (g\sigma)^2 \ln\left(\frac{ p_F+\mu_q}{g\sigma}\right)
	\right] \right\}\ .
\end{equation}
We now employ the local density approximation,
so the local Fermi momentum, $p_F(\bfr)$, is related to the local density,
$\rho_q(\bfr)\equiv\langle\psi^+ \psi\rangle$, by
$\rho_q=(N_cN_f/3\pi^2)p_F^3(\bfr)$.
Furthermore, the local scalar density
$\rho_s(\bfr)\equiv\langle{\overline\psi}\psi\rangle$ is given by
\begin{equation}
\rho_s(\bfr)\ =
2N_cN_f\int\frac {d^3 p}{(2\pi)^3}\frac{M(\bfr)}{\sqrt{p^2+M(\bfr)^2}}\ =\
\frac{N_c N_f}{2\pi^2}  g\sigma(\bfr)\left[\mu_q p_F(\bfr)-g^2\sigma(\bfr)^2
	\ln\left(\frac{p_F(\bfr)+\mu_q}{g\sigma(\bfr)}\right)\right]\ ,
\end{equation}
where we have used that the local Fermi energy 
$E_F(\bfr)=\sqrt{p_F(\bfr)^2+M(\bfr)^2}$ equals the (constant)
chemical potential $\mu_q=\mu/3$.
We note that
\begin{equation}
\left[ \partial_\sigma \Omega_{\rm LSM}(\sigma,\boldsymbol{0};0,\mu)
	\right]_{\sigma=\sigma(\bfr)} \
=\ U^\prime(\sigma(\bfr)) + g \rho_s(\bfr)\ .
\label{bubble2}
\end{equation}
Then the stationary Euler-Lagrange equation with $\piv=\boldsymbol{0}$ 
and ${\overline\psi}\psi$ replaced by $\rho_s(\bfr)$,  
provides an equation for the local value
 of the order parameter, $\sigma(\bfr)$,
\begin{equation}
	\boldsymbol{\nabla}^2\sigma(\bfr)
	-[U^\prime(\sigma(\bfr))+g\rho_s(\bfr)]\ =\ 0\ .
\label{motion}
\end{equation}
For semi-infinite geometry, the profile of the order parameter, $\sigma(z)$, 
can then be obtained by solving the corresponding Euler-Lagrange equation,
\begin{equation}
\partial_z^2\sigma(z) - \left[\partial_\sigma
	\Omega_{\rm LSM}(\sigma)\right]_{\sigma=\sigma(z)}\ =\ 0\ ,
\label{bubble1}
\end{equation}
with the boundary conditions that the order parameter approach the
zero-temperature coexistence values far from the surface,
\begin{equation}
\sigma(z\to-\infty)\ \to\ \sigma_1(T=0)=f_\pi\ ,\,\,\
\sigma(z\to+\infty)\ \to\ \sigma_2(T=0)=0.13\,f_\pi\ .
\end{equation}

\begin{figure}[tbh]	
\vspace{0.5cm} 
\epsfig{figure=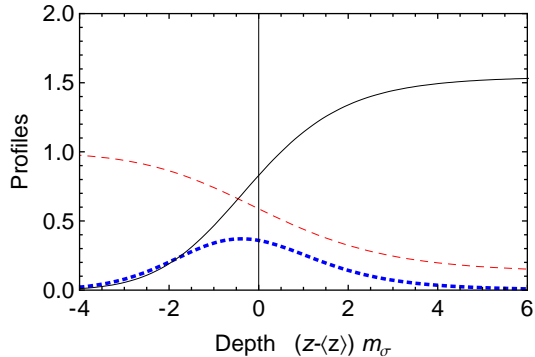,angle=0,width=7cm}
\caption{The order parameter $\sigma(z)$ in units of $f_\pi$ (dashed), 
the net baryon density $\rho(z)$ in units of $\rho_0$ (solid),
and the surface location function $g(z)$ (dotted),
as functions of the dimensionless depth $(z-\langle z \rangle) m_\sigma$
relative to the average surface location.}
\label{fig1}
\end{figure}

Once $\sigma(z)$ is known, so is the mass $M(z)$,
and we can then obtain the local Fermi momentum $p_F(z)$ and,
consequently, the local net baryon density $\rho(z)$. 
One may then define the associated interface
location function \cite{RandrupPRC79,MyersBook},
\begin{equation}
 g(z) \equiv \frac{\partial_z  \rho(z)}{\rho_2 -\rho_1}\ ,
\end{equation}
which allows us to obtain additional information,
such as the mean interface location, 
${\bar z}=\langle z\rangle\equiv\int zg(z)dz$, 
the associated interface width $b$, 
where $b^2=\langle (z-{\bar z})^2 \rangle$, 
as well as a measure of the profile skewness 
which is given by the dimensionless parameter 
$\gamma_3 \equiv \langle (z- {\bar z})^3 \rangle/b^3$ \cite{MyersBook}.
The calculated results for $\sigma(z)$, $\rho(z)$, and $g(z)$ 
are shown in Fig.\ \ref {fig1}.  
The origin is conveniently located at $\langle z \rangle$,
the width is $b=1.85\,a$ (where $a=1/m_\sigma\approx0.33\,\fm$)
and the skewness is $\gamma_3=0.6$ .
As the temperature is increased,
the profiles widen progressively and grow more symmetric,
as also found in Ref.\ \cite{RandrupPRC79}.

\subsection{The Nambu--Jona-Lasino model}

We now consider the standard version of the two-flavor NJL model \cite{njl}. 
Its Lagrangian density is based
on a chirally symmetric four-fermion interaction,
\begin{equation}
\mathcal{L}_{\rm NJL}={\bar \psi}\left(i{\gamma }_{\mu }{\partial }^{\mu
  } -m
  \right) \psi+ G \left[ ({\bar \psi}\psi)^{2}+({\bar{\psi}} i\gamma
  _{5}{\vec \tau}\psi )^{2}\right] ,  
\label{njl2}
\end{equation}
where $\psi$ is to be interpreted as in the LSM.
Furthermore, it is assumed that $m_u=m_d$
so the mass matrix is given by $m_c= m{\rm diag}(1,1)$. 
In the mean-field approximation,
the grand canonical potential reads \cite{scavenius, sasaki,prc1}
\begin{equation}
\Omega_{\rm NJL}(\mu,T)\ =\ \frac{(M-m)^2}{4G}
-2N_f N_c \int_{p < \Lambda}\frac{d^{3}\boldsymbol{p}}{( 2\pi)^3}
\left\{E-T\ln \left[1-n^+\right]-T\ln\left[1-n^-\right]\right\}\ ,
\label{omeganjl}
\end{equation}
with the same definitions as used in the LSM. 
For each value of $T$ and $\mu$ the dynamical mass is of the form
$M{\rm diag}(1,1)$ because of the assumption of isospin symmetry 
($m_u=m_d=m$) and chemical equilibrium ($\mu_u=\mu_d=\mu$);
alternative scenarios may also be considered \cite {buballa}.
The single dynamical mass $M$ is then obtained by minimizing $\Omega$ 
with respect to $M$, leading to the well known gap equation,
\begin{equation}
M\ =\ m + 2G N_c N_f \int_{p<\Lambda} 
\frac{d^3\boldsymbol{p}}{(2\pi)^3}\frac {M}{E} \left[1-n^+-n^-\right]\ .
\label{gap}
\end{equation}

Although the thermodynamic potentials for LSM (Eq.\ (\ref{omegalsm}))
and the NJL model (Eq.\ \ref{omeganjl}))
have the same structure as far as the loop contribution is concerned,
some important differences between the two models exist. 
First, we note  that within the NJL
the quark mass acquires its constituent value only when quantum
corrections (loop terms) are computed. Therefore, contrary to the LSM,
the divergent term represented by the second term on the right-hand
side of Eq.\ (\ref{omeganjl}) plays a central role regarding the
(dynamical) chiral symmetry breaking. Another difference between the
LSM and the NJL model, in 3+1 dimensions, is that the latter is
not renormalizable since the coupling $G$  carries dimensions ($\rm
energy^{-2}$). This means that potential divergencies cannot be
systematically eliminated by a redefinition of the original
parameters. By considering it as  an effective model, one gives up the
very high energies and evaluates all  the integrals up to an
ultraviolet (non-covariant) cutoff $\Lambda$, as the notation in
Eqs.\ (\ref {omeganjl}) and (\ref {gap}) implies. Then $\Lambda$ is
treated as a ``parameter'' which will be fixed, together with $G$ and
$m$, so as to yield the values of physical observables such as
$m_\pi$, $f_\pi$ that reproduce the phenomenological value of 
$\langle{\overline\psi}\psi\rangle$. 
For example, in Ref.\ \cite {scavenius}
the authors reproduce $f_\pi=93\,\MeV$ and $m_\pi=138\,\MeV$
using $\Lambda=631\,\MeV$ and $G\Lambda^2= 2.19$ with $m=5.5\,\MeV$. 
These parameter values, which we label ``set I'', 
predict a first-order phase transition starting at $T=0, \mu=1045.5\,\MeV$ 
and ending at the critical point $(T_c=46\,\MeV,\mu_c= 996\,\rm MeV)$,
while the constituent quark mass in vacuum is $M^{\rm vac}= 337\,\rm MeV$. 
In their study of the chiral phase transition 
in the presence of spinodal decomposition the authors in Ref.\ \cite {sasaki}
use $\Lambda=587.9\,\rm MeV$ and $G\Lambda^2=2.44$ with $m=5.6\,\rm MeV$ 
in order to reproduce $f_\pi=92.4\,\MeV$ and $m_\pi= 135\,\MeV$ 
obtaining $M^{\rm vac}= 400\,\MeV$. 
We shall also consider these parameter values,  which we label ``set II'', 
in order to estimate the influence of different parametrizations 
in the estimation of $\gamma_T$.
 With parameter set II the first-order transition line starts at 
$(T=0,\mu=1146.3\,\MeV)$ and ends at $(T_c=81\,\MeV,\mu_c=990\,\MeV)$.
In general, a larger value of $G \Lambda^2$ enlarges the coexistence region.
As in the LSM case, the quantities $\rho_1$, $\rho_2$, $\rho_g$, ${\cal E}_g$,
 and $\Delta f(\rho)$ entering the expression (\ref{gamma})
for the surface tension can be obtained from the equation of state. 
As  already emphasized,the numerical value for the length scale $a$
is chosen to be $1/m_\sigma \simeq 0.33\,\fm$  
(which is about the value found in a Thomas-Fermi application 
to the NJL model \cite{russos}). 
The remaining two numerical inputs, $\rho_g$ and $\epsilon_g$, 
are automatically fixed once the EoS has been determined.

\begin{figure}[tbh]	
\vspace{0.5cm} 
\epsfig{figure=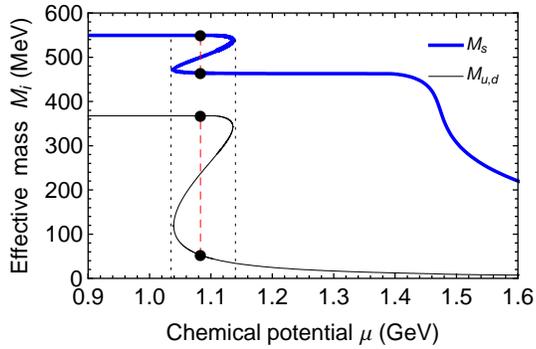,angle=0,width=7cm}
\caption{The constituent quark masses as functions of of the baryon chemical
potential $\mu$ for $T=0$ in the three-flavor NJL model.
The top curve shows $M_s$, while the curve represents $M_u=M_d$. 
Phase coexistence occurs at $\mu=1083\,\MeV$ and the corresponding
mass values are indicated by the dots.  These are are joined by the
Maxwell lines (dashed) that trace out the gradual phase conversion
when full equilibrium is maintained;
$M_s$  then decreases from 549 to 464.4 $\MeV$,
while $M_u=M_d$ decreases from 367.6 MeV to 52.5 $\MeV$.
Between the two dotted lines there are three solutions for a given $\mu$:
stable matter, metastable matter, 
and (between the two extrema) spinodally unstable matter.}
\label{fig3}
\end{figure}		

\begin{figure}[tbh]	
\vspace{0.5cm} 
\epsfig{figure=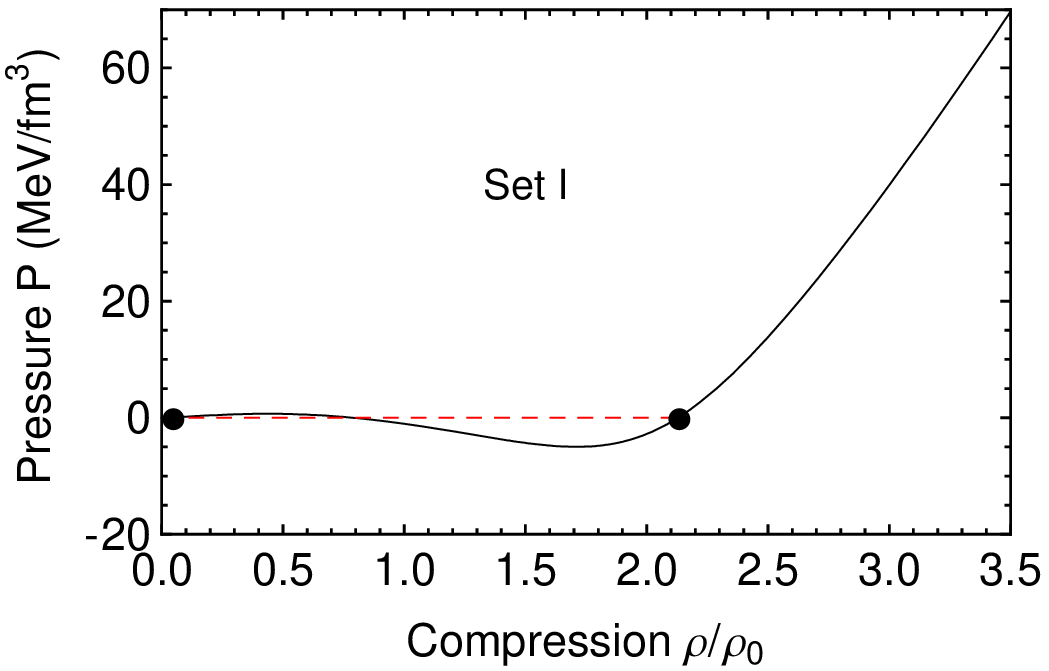,angle=0,width=7cm}
\epsfig{figure=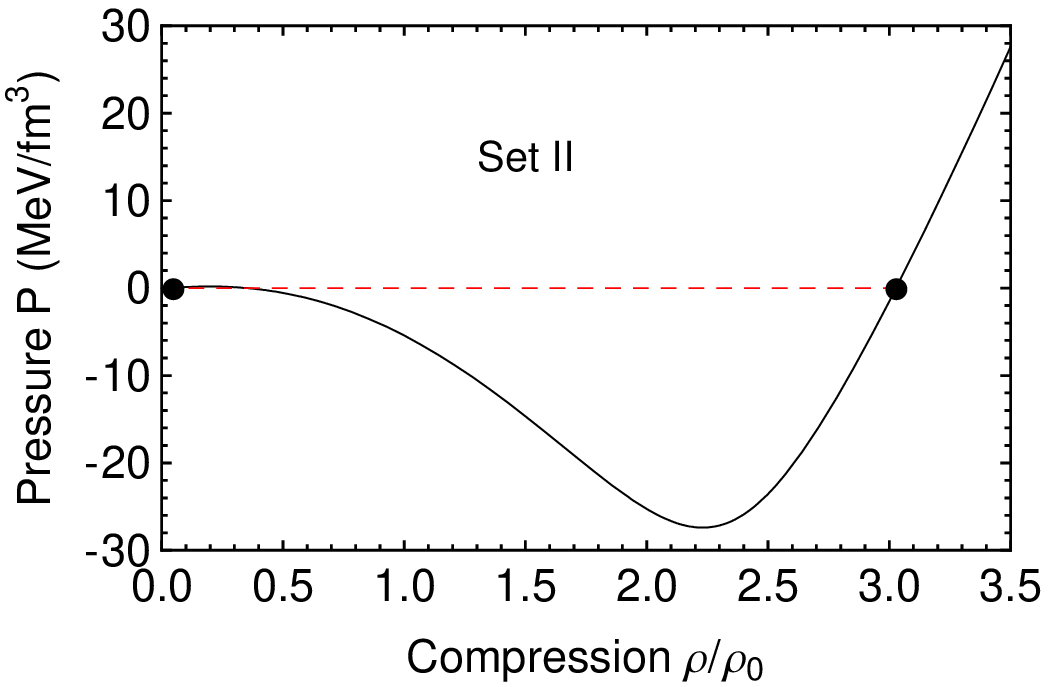,angle=0,width=7cm}
\caption{ The pressure, $P$, 
as a function of the compression $\rho/\rho_0$ for the NJL model
as obtained with the two parameter sets. 
The two coexistence points (dots) are joined by the Maxwell line (dashed)
along which the global equilibrium evolves as the density is increased
through the phase coexistence region.}
\label{fig4}
\end{figure}		

\begin{figure}[tbh]	
\vspace{0.5cm} 
\epsfig{figure=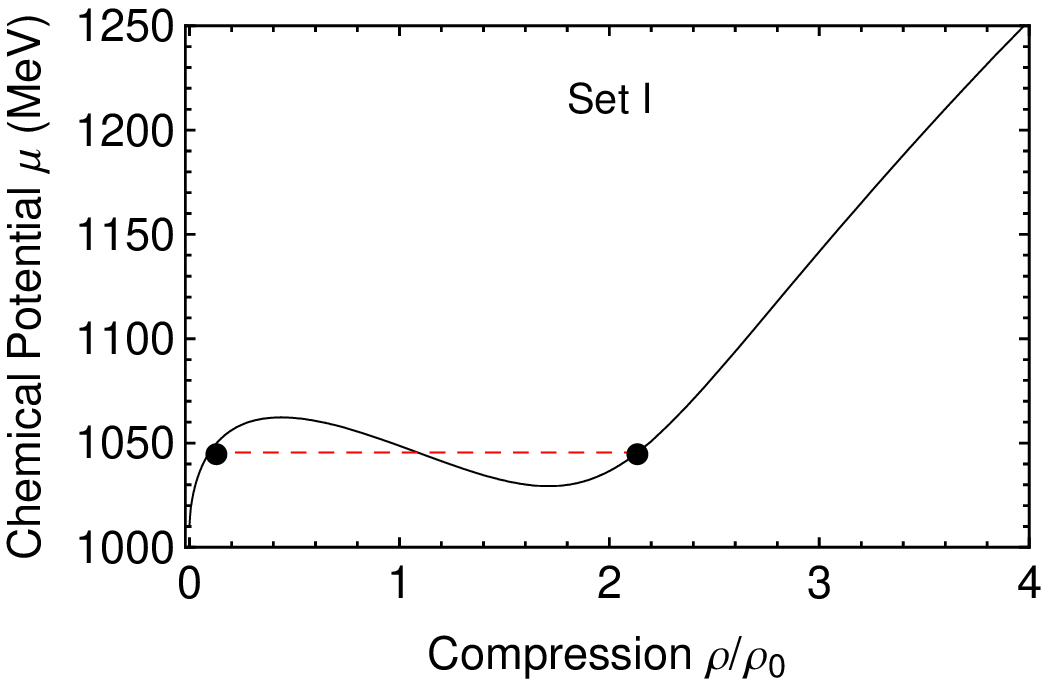,angle=0,width=7cm}
\epsfig{figure=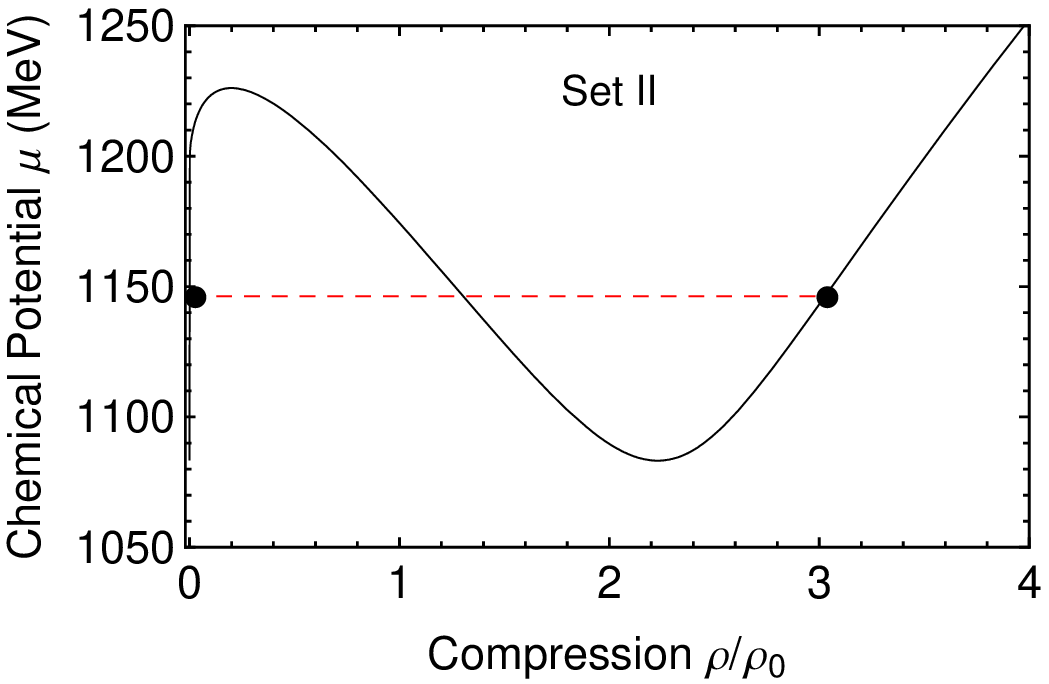,angle=0,width=7cm}
\caption{The baryon chemical potential, $\mu$, as a function of the
compression $\rho/\rho_0$ for the NJL model
as obtained with the two parameter sets. 
The two coexistence points (dots) are joined by the Maxwell line (dashed)
along which the global equilibrium evolves
through the phase coexistence region.}
\label{fig5}
\end{figure}		

\begin{figure}[tbh]	
\vspace{0.5cm} 
\epsfig{figure=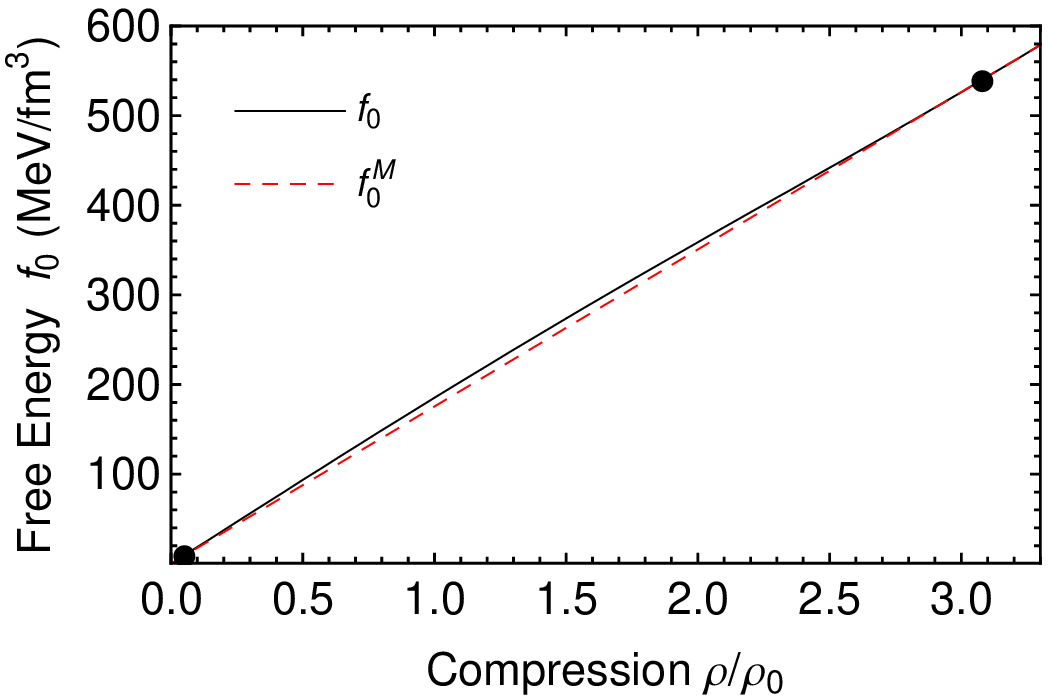,angle=0,width=7cm}
\epsfig{figure=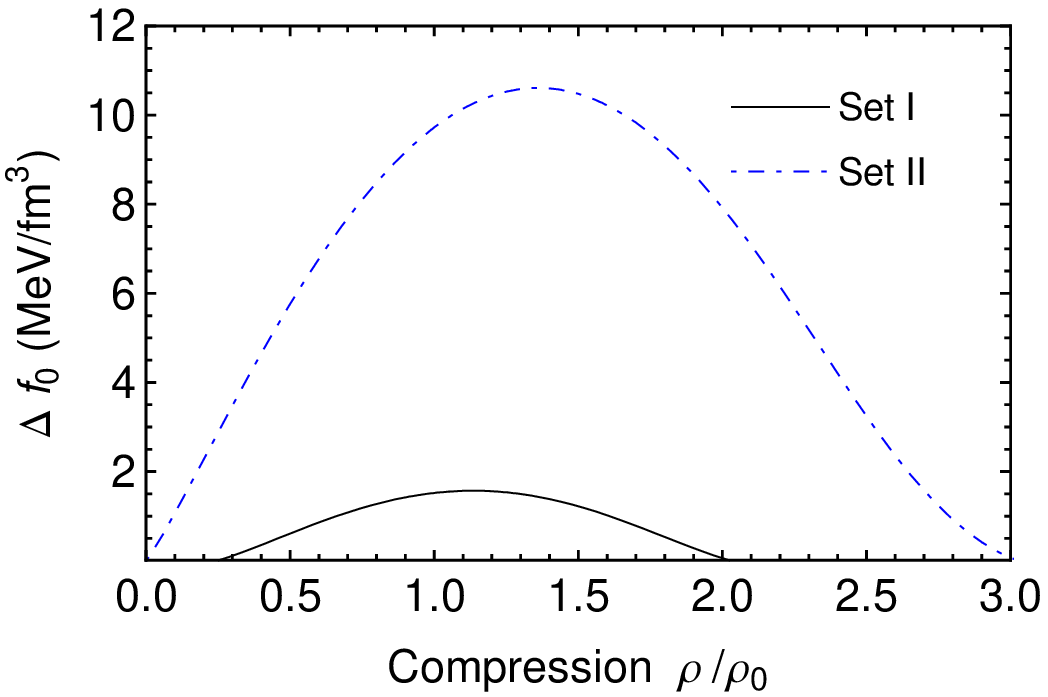,angle=0,width=7cm}
\caption{Left panel: The free energy density at zero temperature
$f_0(\rho)$ (solid) and the associated Maxwell line $f_0^M(\rho)$ (dashed)
as functions of the compression $\rho/\rho_0$ 
for the NJL model with parameter set II. 
The Maxwell line is tangent to $f_0(\rho)$ at the two coexistence 
densities $\rho_1=0$ and $\rho_2=3.027\,\rho_0$. 
Right panel: 
The quantity $\Delta f_0(\rho)$ as a function of $\rho/\rho_0$
obtained with the NJL model for parameter set I (continuous) 
and parameter set II (dot-dashed).}
\label{fig6}
\end{figure}		

\begin{figure}[tbh]	
\vspace{0.5cm} 
\epsfig{figure=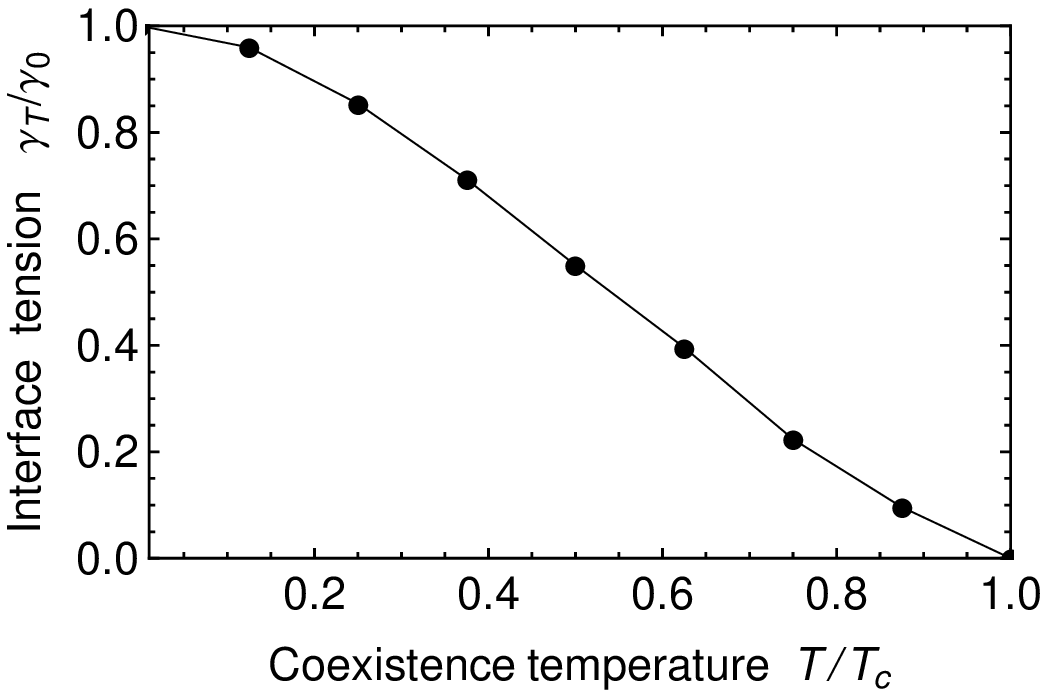,angle=0,width=7cm}
\caption{The surface tension $\gamma_T$
(relative to its zero-temperature value $\gamma_0$)
as a function of the temperature $T$
(measured relative its critical value $T_c$),
as obtained with the NJL model for parameter set II.}
\label{fig7}
\end{figure}		

\section{The EoS for the NJL model with three quark flavors}
\label{NJL3}

In stellar modeling, the structure of the star depends on the
assumed EoS built with appropriate models while the true
ground state  of matter remains a source of speculation. 
It  has been argued \cite{itoh,Witten,Haensel,Alcock} that 
{\em strange quark matter} (SQM) is the true ground  state of all matter 
and this hypothesis is known as the Bodmer-Witten conjecture. 
Hence, the interior of neutron stars should be composed predominantly of
$u,d,s$ quarks (plus leptons if one wants to ensure charge neutrality
 which is not the case in the present work). 
Strangeness is implemented in the $SU(3)$ version of the NJL model 
which is given by 
\begin{equation}
\mathcal{L}\ =\ {\bar\psi}(i\gamma_\mu\partial^\mu-m)\psi
+G\sum_{a=0}^8 \left[({\bar\psi}\lambda^a\psi)^2
	+(\bar{\psi}i\gamma_5\lambda^a\psi)^2\right] 
-K \left\{{\rm det}_f[\bar\psi(1+\gamma_5)\psi] 
	 +{\rm det}_f[\bar\psi(1-\gamma_5)\psi]\right\}\ ,
\label{njl3}
\end{equation}
where $\psi=(u,d,s)^T$ denotes a quark field with three flavors 
(and three colors), 
and $m= {\rm diag}_f(m_u,m_d,m_s)$ is the corresponding mass matrix. 
Here we assume $m_u=m_d \ne m_s$ indicating that isospin symmetry is observed 
while the $SU(3)$ flavor symmetry is explicitely broken. 
The eight Gell-Mann matrices are represented by $\lambda^a$ ($a=1,...,8$) 
and $\lambda^0=\sqrt{2/3}\,\boldsymbol{I}$. 
More details concering this version of the NJL model
can be found in Ref.\ \cite {buballa}. 
In the mean-field approximation the thermodynamical potential is given by
\begin{equation}
 \Omega_{{\rm NJL}_3}(T,\mu_q)\ =\  \sum_{q=u,d,s} 
\left[\Omega_{M_q}(T,\mu_q)+2G\phi^2_q\right]-4K\phi_u\phi_d\phi_s\ .
\end{equation}
The term $\Omega_{M_q}$, 
which represents the contribution of a gas of quasiparticles with mass $M_q$, 
is given by
\begin{equation}\label{omeganjl3}
\Omega_{M_q}\ =\ -2N_c \int_{p<\Lambda}\frac{d^{3}\boldsymbol{p}}{( 2\pi)^3}
\left\{E_q-T\ln\left[1-n_q^+ \right]-T\ln\left[1-n_q^-\right]\right\}\ ,
\end{equation}
where $E_q^{2}=\boldsymbol{p}^2+M_q^{2}$
and $n^\pm_q = \{1+\exp[( E_q\mp\mu_q)/T]\}^{-1}$
represent the particle/antiparticle distribution function. 
For the quark condensates, $\phi_q=\langle{\bar\psi}_q\psi_q\rangle$, one has
\begin{equation}
\phi_q\ =\ -2 N_c  \int_{p< \Lambda}\frac{d^3\boldsymbol{p}}{(2\pi)^3} 
\frac {M_q}{E_q}\left[1-n_q^+-n_q^-\right]\ .
\label{gap3}
\end{equation}
Finally, the gap equation is
\begin{equation}
M_i\ =\ m_i-4G \phi_i+2K\phi_j\phi_k\ ,\
(i,j,k)={\rm any \,\,\,permutation\,\,\, of}\,\,\, (u,d,s)\ ,
\end{equation}
which contains a non-flavor mixing term proportional to $G$ 
as well as a flavor mixing term proportional to $K$. 
In our numerical analysis we adopt the parameter values of Ref.\ \cite{hufner}
which are $m_u=m_d=5.5\,\MeV$, $m_s=140.7\,\MeV$, $G\Lambda^2=1.835$, 
$K \Lambda^5=12.36$, and $\Lambda= 602.3\,\MeV$. 
Then, at $T=0$ and $\mu_f=0$, one reproduces $f_\pi=92.4\,\MeV$, 
$m_\pi=135\,\MeV$, $m_K=497.7\,\MeV$, and $m_{\eta^\prime}=960.8\,\MeV$. 
For the quark condensates one obtains $\phi_u=\phi_d=-(241.9\,\MeV)^3$, 
and $\phi_s=- (257.7\,\MeV)^3$. 
The constituent quark masses are then given by $M_u=M_d=367.7\,\MeV$
and $M_s=549.5\,\MeV$. 
The  pressure, $P$, and energy density, $\cal E$, follow from
the usual expressions,
\begin{equation}
 P=-\Omega_{{\rm NJL}_3}(T,\{\mu_q\}) \,\,\,\,\,\,\,\,\,{\rm and}
\,\,\,\,\,\,\,\,\,P = Ts-{\cal E} + \sum_{q=u,d,s} \mu_q \rho_q \,\,.
\end{equation}

Here, for simplicity,
we take the chemical equilibrium condition, $\mu_u=\mu_d=\mu_s=\mu_q=\mu/3$
which yields $M_u=M_d$ also at finite $T$ and/or $\mu$. 
Of course, for a realistic description of neutron-star matter,
charge and strangeness neutrality need to be taken into account, 
which is technically straightforward.

The phase diagram for this three-flavor quark model model in the 
$\mu$--$T$ and $\rho$--$T$ planes can be found in Ref.\ \cite {pedro};  
the first-order transition line starts at ($T=0$, $\mu=1083\,\MeV$) 
and ends at the critical point ($T_c = 67.7\,\MeV$, $\mu_c=955.2\,\MeV$).  
As already discussed, 
we need the EoS inside the phase coexistence region which can be obtained 
by examining how the effective masses behave in this domain. 
This behavior is shown in  Fig.\ \ref {fig3} for $T=0$;
these results go beyond those of Ref.\ \cite {sasaki} 
by also considering the strange quark mass. 
To understand this figure, let us recall that, in most situations, 
one is generally interested only in those solutions of the gap equation 
that correspond to global (stable) minima of the thermodynamical potential.
However, when a first-order phase transition is present,
there are two different such solutions for the same thermodynamic conditions
of temperature, chemical potential, and pressure 
(corresponding to the solid dots on Fig.\ \ref{fig3}).  
As the net baryon density (which serves as a convenient order parameter)
is increased from its lower coexistence value $\rho_1$ to its higher
coexistence value $\rho_2$, the thermodynamically favored state is a
Maxwell mixture of the two coexisting phases and the overall average
of the energy per net baryon or the effective mass, for example,
evolve monotonically along the socalled Maxwell line,
as the composition of the mixture changes from being entirely one phase
to being entirely the other.
In the region between the dotted lines the gap equation has three solutions,
leading to the back-bending evolution brought in that diagram.
It is  precisely this typical first-order behavior 
that will be reflected in the thermodynamical quantities,
such as the pressure and densities, as Figs.\ \ref{fig4} and \ref{fig5} show.  
This behavior is responsible for the fact that there is a (positive)
deviation $\Delta f(\rho)$,
which then in turn leads to the surface tension.

\section{Numerical Results}
\label{results}

We now turn to our numerical results for the surface tension $\gamma_T$. 
To this end we need to determine the free energy density $f_T(\rho)$,
which requires the evaluation of $P_T(\rho)$ and $\mu_T(\rho)$
for uniform matter thermodynamically unstable region of the phase diagram.
For the considered temperature $T$, the associated density region
is bounded by the two coexistence densities $\rho_1$ and $\rho_2$,
for which the chemical potential $\mu$ has the same value,
as does the pressure $P$.
As the density $\rho$ is increased through the lower mechanically metastable
(nucleation) region, $\mu$ and $P$ rise steadily until the lower spinodal
boundary has been reached.
Then, as $\rho$ moves through the mechanically unstable (spinodal) region,
both $\mu$ and $P$ decrease until the higher spinodal boundary is reached.
They then increase again as $\rho$ moves through the higher mechanically
metastable (bubble-formation) region,
until they finally regain their original values at $\rho=\rho_2$.
It is convenient to express the (net) baryonic density $\rho$
in units of the nuclear saturation density, $\rho_0=0.153/\fm^3$.
Generally, as is common practice,
 we subtract from the pressure any finite value it may have in the vaccum.

\subsection{Zero temperature}

Let us start with $T=0$ for which the relevant results can be readily obtained
by taking the $T\to 0$ limit in Eqs.\ (\ref{omegalsm}) and (\ref{omeganjl}) 
(see, {\em e.g.}\ Refs. \cite{leticia,prc1,buballa}). 
Figure \ref {fig4} shows the pressure as a  function of 
the degree of compression $\rho/\rho_0$ obtained with the NJL model 
for both parameter sets I and II;
the latter has a stronger coupling and a larger coexistence region.
A qualitatively similar behavior is observed in Fig.\ \ref {fig5} which
shows $\mu$ as a function of $\rho/\rho_0$ for the NJL model with
both parameter sets. Figure \ref{fig6} shows the behavior of the
free energy $f_0(\rho)$ and its corresponding Maxwell line $f^M_0(\rho)$ 
for parameter set II. 
In the right panel of Fig.~\ref{fig6}
we display the difference between these two free energies,
$\Delta f(\rho)\equiv f_0(\rho)-f^M_0(\rho)$ for both parameter sets;
this is the key quantity for the determination of the surface tension.   
 The LSM and the three-flavor NJL model yield  similar results. 
At temperatures below criticality, $T<T_c$,
the thermodynamical potential has two degenerate minima determining
the densities of the two coexisting phases, $\rho_1$ and $\rho_2$.
In all cases studied here, 
the lower coexistence density vanishes, $\rho_1=0$.
As for the higher coexistence density,
the LSM yields $\rho_2/\rho_0\simeq1.54$,
the two-flavor NJL model yields $\rho_2/\rho_0\simeq2.13$ with set I 
and $\rho_2/\rho_0\simeq3.03$ with set II,
while the three-flavor NJL model gives $\rho_2/\rho_0\simeq2.62$.
It clear from this figure that set I produces a much weaker phase transition 
because $\Delta f(\rho)$  is much smaller that for set II, 
as is indeed reflected in the $\gamma_T$ values shown in Table I. 
In fact, set II produces a greater coexistence region 
($\rho_1\simeq 0$, $\rho_2\simeq 3.03$,$\rho_0,T_c= 81\,\MeV$)
when compared to set I 
($\rho_1\simeq0$, $\rho_2\simeq 2.13\rho_0$, $T_c= 46\,\MeV$),
 which is in accordance with the well known fact that 
set II should cause the size of the first-order transition line
to be longer than the one produced by set I which has a weaker coupling. 
Further refinements, such as finite-$N_c$  corrections \cite{bowman}, 
contributions from thermal flucutations,
and the inclusion of a repulsive vector interaction \cite{FukushimaGV},
also tend to shrink the first-order transition line \cite{prc1} 
so that, within a fixed parameter set, 
one should expect these effects to reduce $\gamma_T$.

Table I summarizes all our results for $\gamma_0$ and also lists
the characteristic values ${\cal E}_g$ and $\rho_g$ 
as well as the location of the critical point $(T_c,\mu_c)$. 
The table also provides information related to thermodynamic potential 
at $T=0$ and $\mu=0$ by showing the values of the constituent quark mass 
in vacuum ($M^{\rm vac}$),
which is related to the distance from the global minimum to the origin, 
as well as the bag constant  which gives the energy difference 
between the local maximum and the global minimum of the potential in vacuum;
these values were taken from Refs.\ \cite {scavenius,buballa}.

\begin{table}[htb]	
\begin{center}
\begin{tabular}{l||rlllllll}
Model &  
$\gamma_0~~ $ & $ ~{\cal B}_0 $& $M^{\rm vac}_{u,d}$ &$M^{\rm vac}_s$&
$T_c  $ &  $\mu_c  $ &$\rho_g/\rho_0$
& ~${\cal E}_g $  \\ \hline
NJL (I)        &  7.11 &  100    & 337 &\,\,\,\,\,--&
46 & 996 &2.00&342.85 \\ 
NJL (II)~            &  ~30.25   &   141.4     &
400 &\,\,\,\,\,--&81 & 990 &  2.42&495 \\ 
LSM            &  13.18   &   ~60     &
306.9 &\,\,\,\,\,--& 99 & 621 &  1.19&219.25  \\ 
${\rm NJL}_3$ & 20.42&291.7&367.6~&549.5&67.7&955.2&1.87&326.8
\\ \hline  
\end{tabular}
\end{center}
\caption{\label{tabfit} Summary of inputs and results. 
The length parameter was taken as $a=0.33\,\fm$.
The zero-temperature bag constant ${\cal B}_0$ and the characteristic energy 
density ${\cal E}_g$ are given in $\MeV/\fm^3$, 
while the effective quark masses in vacuum $M^{\rm vac}_i$
as well as the critical values $\mu_c$ and $T_c$ are in $\MeV$.
The resulting zero-temperature surface tension $\gamma_0$ 
(given in $\MeV/\fm^2$)
may be compared with the value $\gamma_0=12.96\,\MeV/\fm^2$ obtained 
in Ref.\ \cite{leticia}) with the LSM in the thin-wall approximation.}
\end{table}

We finally note that Palhares {\em et al.}\ \cite{leticia},
using the approximation 
$\gamma_T \approx \int|\partial_z\sigma(z)|^2dz$
obtained the estimate $\gamma_0\approx12.98\,\MeV/\fm^2$
which is very close to our LSM value of $13.18\,\MeV/\fm^2$
and also rather similar to the value $12.19\,\MeV/\fm^2$
resulting from evaluating that intergal
using our LSM profile function $\sigma_0(z)$.

\subsection{Finite temperature}

One can easily consider finite temperatures within the employed models.
The interface tension is expected to decrease with increasing temperature 
because both the coexistence densities $\rho_i$
and the associated free energy densities $f_T(\rho_i)$
move closer together at higher $T$;
they ultimately coincide at $T_c$ where, therefore, the tension vanishes. 
This general behavior is confirmed by our calculations,
as shown in Fig.\ \ref {fig7}.
The LSM, the NJL with parameter set I, and the three-flavor NJL
display similar behaviors. 
The temperature dependence of the surface tension may be relevant for
the  thermal formation of quark droplets in cold hadronic matter
found in ``hot'' protoneutron stars whose temperatures, $T_*$, 
are of the order 10--20 MeV \cite{bruno,hotstars,Olesen}. 
For $T_*$ the relevant value of $\gamma_{T_*}$ 
may be estimated by using table I together with Fig.\ \ref {fig6}.
For example, the three-flavor NJL model yields
$\gamma_{T_*}\approx 14-18\,\MeV/\fm^2$.
The temperature dependence of the surface tension is also
important in the context of heavy-ion collisions, 
because it determines the favored size of the clumping
caused by the action of spinodal instabilities 
as the expanding matter traverses the unstable phase-coexistence region.

\section{Conclusions}
\label{conclude}

In this work we have shown that the interface tension related to a first-order
phase transition may be evaluated once the uniform-matter equation of state
is available for the unstable regions of the phase diagram.
It is a convenient feature of the method employed
that knowledge of the interface profile functions is is not required, 
because their determination can be quite complicated, 
as is the case for NJL model \cite{russos} (although it is easy for the LSM).
In addition to the EoS,
the geometrical approach also requires a proper setting of three input 
parameters, namely the characteristic densities $\rho_g$ and ${\cal E}_g$
together with the length scale $a$.
While this does encumber the numerical results with some degree of uncertainty,
our zero-temperature LSM result, $\gamma_0=13.18\,\MeV/\fm^2$, agrees within 
a few percent with the approximate value obtained in Ref.\ \cite{leticia},
thus suggesting that those parameters were chosen reasonably.

The surface tension determined in the present fashion is entirely
consistent with the employed model, 
including the approximations and parametrizations adopted. 
For the effective quark models employed here,
this amounts to considering {\it all} the solutions to the gap equation
(stable, metastable and unstable) 
and determine the relevant effective quark masses.
In most non-perturbative approximations 
(large $N_c$, mean field, {\em etc.}) 
the various quantities of interest, such as the free energy density, 
become functions of this effective mass and will therefore also reflect 
the metastable and unstable character of the configuration considered.
As a cross check on our procedure, we have evaluated $\gamma_0$ 
for the LSM obtaining a result that differs by only about $2\%$ 
from estimates based on the thin-wall approximation \cite{leticia}.  
We have investigated the two-flavor NJL model 
as well as its more realistic three-flavor version.

Our main conclusion is that all these effective models generate relatively
low values for the the surface tension.
This would favor the formation of quark matter and may thus have important 
astrophysical consequences regarding the existence of pure quark stars. 
Of particular interest is the three-flavor NJL result, 
$\gamma_0=20.34\,\MeV/\fm^2$,
because this model is widely used in studies related to neutron stars. 
Here, for simplicity, we have considered pure quark matter 
where all flavors share the same chemical potential,
but it is just a technical matter to generalize our procedure 
so as to include leptons ($e,\mu$) in order to enforce $\beta$ equilibrium
($\mu_d=\mu_s=\mu_u +\mu_e \, , \, \mu_e=\mu_\mu$),
although the additional chemical potential introduces an increased
degree of complexity into the features of the phase transition.

In principle, more refined treatments, such as the Polyakov-NJL model, 
can also be considered within the same framework. 
However, because the effects of the Polyakov loop become more important 
above $100\,\MeV$ \cite{costa} we believe that our results, 
especially the three-flavor ones, can be considered as reasonably accurate,
although numerical variations may arise 
due to the parametrizations and approximations adopted.

\acknowledgments

MBP was partially supported by CNPq-Brazil,
while VK and JR were supported by the Office of Nuclear Physics 
in the US Department of Energy's Office of Science under Contract No.
DE-AC02-05CH11231.
We thank Eduardo Fraga, Sidney Avancini, and D\'{e}bora Menezes 
for their comments and suggestions.
VK would like to thank the Departamento de F{\i}sica, Universidade Federal de
 Santa Catarina for the kind hospitality during the completion of this work.


\end{document}